\documentclass{SCIS2016}
\usepackage{hyperref}
\usepackage{graphics}
\usepackage{caption}

\newtheorem{mydef}{Definition}

\newtheorem{conj}{Conjecture}
\begin{document}

\ArticleType{RESEARCH PAPER}{}
\Year{Xxxx}
\Month{Xxxxxx}
\Vol{Xx}
\No{X}
\DOI{xxxxxxxxxxxxxx}
\ArtNo{xxxxxx}
\ReceiveDate{Xxxxxxx x, xxxx}
\AcceptDate{Xxxxxxx x, xxxx}
\OnlineDate{Xxxxxxx x, xxxx}

\title{JS-MA: A Jensen-Shannon Divergence Based Method for Mapping Genome-wide Associations on Multiple Diseases}{JS-MA: A Jensen-Shannon Divergence Based Method for Mapping Genome-wide Associations on Multiple Diseases}

\author[1]{Xuan GUO}{{xuan.guo@unt.edu}}

\AuthorMark{Xuan Guo}

\AuthorCitation{Xuan Guo}

\address[1]{Department of Computer Science and Engineering, University of North Texas, Texas {\rm 76203}, USA}
\maketitle

\abstract{Taking advantages of high-throughput genotyping technology of single nucleotide polymorphism (SNP), large genome-wide association studies (GWASs) have been considered as the promise to unravel the complex relationships between genotypes and phenotypes, in particularly common diseases. However, current multi-locus-based methods are insufficient, in terms of computational cost and discrimination power, to detect statistically significant interactions and they are lacking in the ability of finding diverse genetic effects on multifarious diseases. Especially, multiple statistic tests for high-order epistasis ($ \geq $ 2 SNPs) will raise huge analytical challenges because the computational cost increases exponentially as the growth of the cardinality of SNPs in an epistatic module. In this paper, we develop a simple, fast and powerful method, named JS-MA, using the Jensen-Shannon divergence and a high-dimensional $ k $-mean clustering algorithm for mapping the genome-wide multi-locus epistatic interactions on multiple diseases. Compared with some state-of-the-art association mapping tools, our method is demonstrated to be more powerful and efficient from the experimental results on the systematical simulations. We also applied JS-MA to the GWAS datasets from WTCCC for two common diseases, i.e.  Rheumatoid Arthritis and Type 1 Diabetes. JS-MA not only confirms some recently reported biologically meaningful associations but also identifies some novel findings. Therefore, we believe that our method is suitable and efficient for the full-scale analysis of multi-disease-related interactions in the large GWASs.}


\citationline

\section{Introduction}
Genome-wide association study (GWAS) has been proved to be a powerful tool to identify genetic susceptibility of associations between a trait of interests using statistical tests~\cite{cai3}. Genotype-phenotype association studies have confirmed that single nucleotide polymorphisms (SNPs) are associated with a variety of common diseases~\cite{introMeaning}. The current primary analysis paradigm for GWAS is dominated by the analysis on susceptibility of individual SNPs to one disease a time, which might only explain a small part of genetic causal effects and relations for multiple complex diseases~\cite{introHe}. The word, epistasis, is defined generally as the interaction among different genes~\cite{introEps2}. Many studies have demonstrated that the epistasis is an important contributor to genetic variation in complex diseases. Most common diseases, such as obesity~\cite{introEps3}, cancer~\cite{introEpsBreast}, diabetes~\cite{cai2}, and heart disease~\cite{introEpsHeart}, are complex traits resulting from the joint effects of various genetic variants, environmental factors or their interactions. It is of great interest to identify the genetic risk factors for complex diseases, to understand disease mechanisms, develop effective treatments and improve public health. The cost of genomic technologies is falling exponentially over time. For instance, the Human Genome Project took 13 years and cost \$2.7 billion, whereas the current cost of sequencing a genome is approaching \$1000 and takes less than a week. With the availability of large-scale genotyping technologies together with their rapid improvement, the cost of genome-wide analyses has been widely decreased, and a great number of large-scale genetic association studies is initiated. Complex diseases do not show the ``pure'' inheritance pattern observed in Mendelian diseases, where alterations in a single gene or a unique locus are causal for a phenotype. In a complex disease, multiple genes are involved, each with low-penetrance, where each gene modestly increases the probability of disease and does not ultimately determine disease status. These factors often render the traditional genetic dissection approaches, such as linkage analysis, ineffective tools to study complex diseases. In this article, we consider epistatic interactions as the statistically significant associations of $ d $-SNP modules ($ d \geq 2 $) with multiple phenotypes~\cite{resultReview}.

The problem of detecting high-order genome-wide epistatic interaction for case-control data has attracted more research interests recently. Generally, there are two challenges in mapping genome-wide associations for multiple diseases on a large GWAS dataset~\cite{DCHE}: the first is arose from the heavy computational burden, \textit{i.e.} the number of association patterns increases exponentially as the order of interaction goes up. For example, there are around $ 6.25\times 10^{11} $ statistical tests required to detect pairwise interactions for a moderate dataset with \~~500,000 SNPs. The second challenge is that existing approaches do not have enough statistical powers to report significant high-order multi-locus interaction on multiple diseases. Because of the huge number of hypotheses and the limited sample size, a large proportion of significant associations are expected to be false positives. In recent, many computational algorithms have been proposed to overcome the above difficulties. They can be broadly classified to three categories~\cite{introFourCate}: exhaustive search, stepwise search and heuristics approaches. The naive solution to tack the problem is exhaustive search using statistical tests, like $ \chi^{2} $ test, exact likelihood ratio test or entropy-based test, for all SNP modules~\cite{Wan2}\cite{comment2:02:background}\cite{comment2:01:background}. In order to minimize the huge computation requests, stepwise search strategies select a subset of SNPs or their combinations based on some low-order measurement tests, then extend them to higher order interactions if it is statistically possible~\cite{introNature}\cite{introStep2}. Heuristic methods adopt machine learning or stochastic procedures to search the space of interactions rather than explicitly enumerating all combinations of SNPs~\cite{introRuler}\cite{introBEAM}. More details about the popular GWAS mapping tools can be found in a recent survey~\cite{mesurvey}. To the best of our knowledge, current epistasis detecting tools are only capable of identifying interactions on GWAS data with two groups, \textit{i.e.} case-control study groups. Thus, they are incompetent to discover genetic factors with diverse effects on multiple diseases. Also, they may lose the benefit of alleviating the deficiency of statistical powers by pooling more diseases samples together. 

To the best of our knowledge, most epistasis detecting tools available now are only capable of identifying interactions on the data of GWAS with two groups, \textit{i.e.} case-control studies. These tools are incompetent to discover the genetic factors with diverse effects on multiple diseases. Moreover, only use limited case samples, they may lose the benefit of alleviating deficiency of statistical powers by pooling different disease samples together. Recently, Guo et al. developed a Bayesian inference based method, named DAM, to detect multi-locus epistatic interactions on multiple diseases~\cite{guo2015dam}. From our experiments, DAM took 3 days to finish the processing of real GWAS datasets with limited significant findings. In this manuscript, we present a heuristic method, named JS-AM, by using Jensen-Shannon divergence and a modified $ k $-mean clustering to filter out a candidate set of SNPs showing potentially diverse effects on multiple phenotypic traits. A stepwise evaluation of association is engaged in JS-MA to further determining the genetic effect types of associations. Systematic experiments on both simulated and real GWAS datasets demonstrate that our method is feasible for identifying multi-locus interaction on GWAS datasets and enriches some novel, significant high-order epistatic interactions with specialities on various diseases. 

\section{Materials and Method}

\subsection{Notation}\label{note}
\begin{figure}[h!]
\centering
\includegraphics[width=0.7\linewidth]{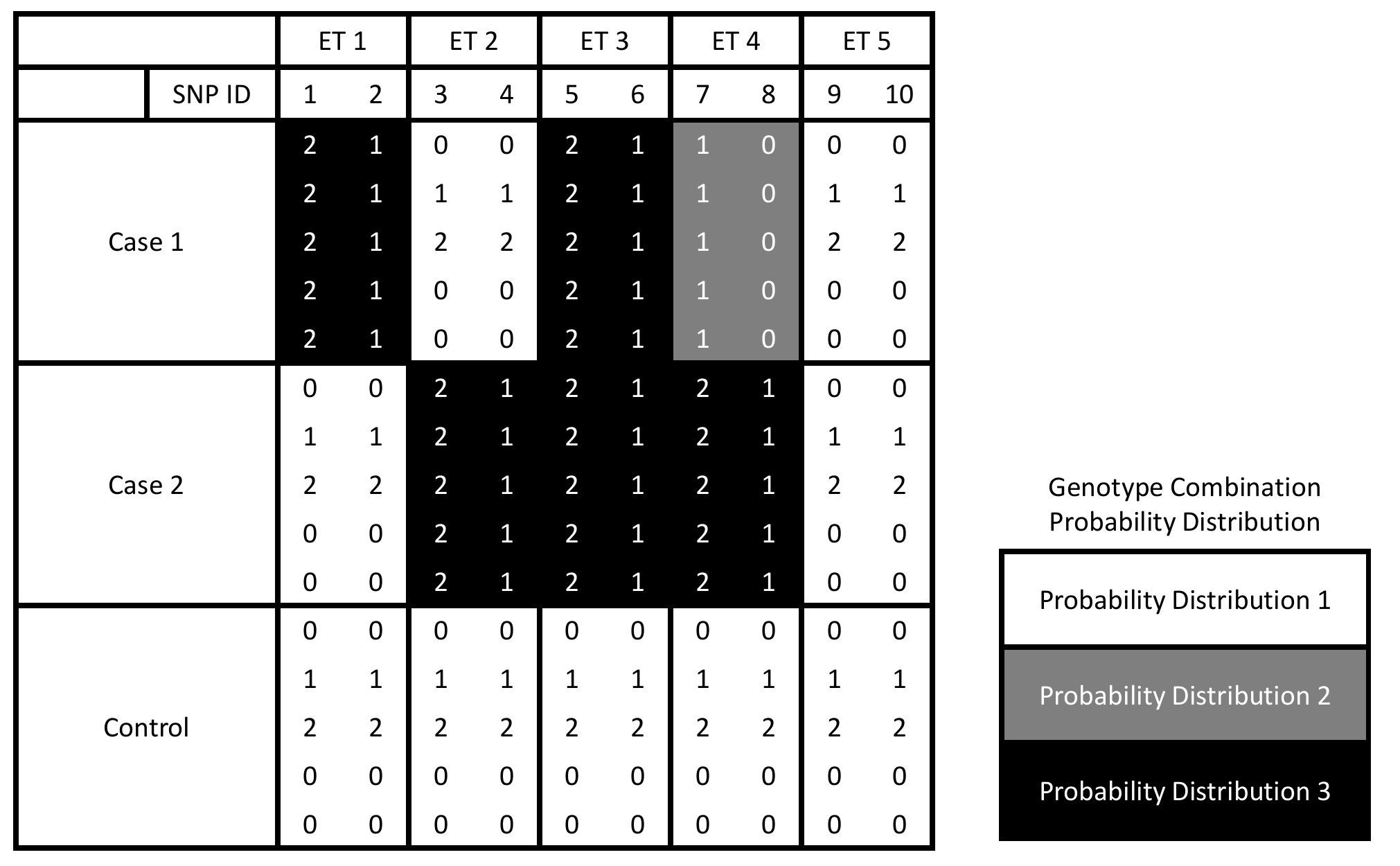}
\caption{The illustration for 5 epistasis types by giving 3 groups. 10 SNPs of ET 1, 2, 3, 4, and 5 are associated with the phenotype traits with interaction between each pair of them.}
\label{fig:group}
\end{figure}

For a GWAS data, let $ L $ denote the total number of groups including $ L-1 $ case groups and one control group, and each group has $ N_{l} $ samples, $ l\in\left\lbrace 1, 2, \ldots, L \right\rbrace  $. Let $ N $ be the total count of samples from $ L $ groups, and $ M $ be the number of diallelic SNP markers in the study. In general, the major alleles are represented as uppercase letters (e.g. $ A $, $ B $,...) and the minor alleles are represented as lowercase letters (e.g. $ a $, $ b $). We use $ \left\lbrace  0, 1, 2 \right\rbrace  $ to represent $ \left\lbrace AA,Aa,aa\right\rbrace  $. $ X $ is utilized to indicate the ordered SNP set where $ x_{i} $ indicates the $ i $-th SNP. Let $ g_{x_{i}, \ldots, x_{j}} $ be the genotype combination vector giving a list of SNPs $ \left\lbrace x_{i}, \ldots, x_{j} \right\rbrace  $. The probability distribution of $ g_{x_{i}, \ldots, x_{j}} $ is denoted as $ p_{g_{x_{i}, \ldots, x_{j}}} $, or $ p_{g} $ for simplicity. The number of effects associated to $ L $ groups is well known as the Bell number~\cite{guo2015dam}. The M SNPs can be assigned to $ B $ different label types, and we call these $ B $ types as Epistasis Types (ET). An example showing all 5 epistasis types for a three-group dataset is in Figure~\ref{fig:group}. In this example, each epistasis including 2 SNPs. We have three different probability distributions of genotype combinations labelled by three colours. We can easily find that SNP 1 and 2 together have contributions to case 1 status, and we call this type effect as ET 1. Similarly, we call SNP 3 and 4, SNP 5 and 6 as ET 2 AND ET 3, respectively. For SNP 7 and 8, the genotype combinations display different effects on two case status and thus have two separate probability distributions. For the last two SNPs, they are related to any case status and show the same probability distribution among three groups. Given $ L $ groups, we denote the number of all combination of two groups as $ H = \left\lbrace h_{1}, \ldots, h_{|H|} \right\rbrace  $, and there are $ L(L-1)/2 $ combinations, which is denoted as $ |H| $. The probability distribution of genotype combinations giving $ h_{i} $ is represented as $ p^{\left( h{i}\right) } $. 

\subsection{Jensen-Shannon Divergence}
We define a distance measurement based on the Jensen-Shannon divergence (JS) for two SNPs. The Jensen-Shannon divergence is a popular distance measurement based on Kullback-Leibler divergence~\cite{sam:jsd}, which evaluate the similarity between two probability distributions. Given two probability distributions, $ p $ and $ q $ with $ g $ categories, the Kullback-Leibler divergence (KL divergence) is defined as follows:

\begin{align}
\mathbb{KL}\left(p \parallel q\right) = \sum_{i=1}^{g}p_{g}log\dfrac{p_{g}}{q_{g}}
\end{align}

The KL divergence is not a distance since it is not symmetric. One symmetric version of the KL divergence is the Jensen-Shannon divergence, defined as follows:

\begin{align}
JS\left( p, q \right) = 0.5\mathbb{KL}\left( p \parallel \frac{p+q}{2}\right)  + 0.5\mathbb{KL}\left(q \parallel \frac{p+q}{2} \right)  
\end{align}

For each group combination $ h_{i} $ and two SNPs, $ x_{i} $ and $ x_{j} $, two probability distributions, $ p^{h_{i}} $ for the first group and $ q^{h_{i}} $ for the second group where each category indicates the possibility of samples. Based on the Jensen-Shannon divergence, we define the distance between two SNPs, $ x_{i} $ and $ x_{j} $ as follows

\begin{eqnarray}
Dist(x_{i}, x_{j}) = \dfrac{\sum_{h_{i}\in H} JS\left( p^{h_{i}}, q^{h_{i}} \right)}{|H|} 
\end{eqnarray} 

Therefore, if these two SNPs do not associated to any case status, the distribution of genotype combination in cases should be similar to the ones in control, and  $ Dist(x_{i}, x_{j}) $ obtains a very small value toward 0; otherwise, $ Dist(x_{i}, x_{j}) $ obtains a large value toward 1.

\subsection{Clustering}

Our goal is to find a list of SNP modules containing $ d $ SNPs, which have larger JS dissimilarities between any two groups than the other modules not in the list. Apparently, it is computational expensive to examine all $ d $ SNP combination when $ d \geq 3 $ and there are millions of SNPs. In order to diminish time complexity, we make use of $ k $-means clustering to group SNPs into clusters where SNPs having jointly effect associating with cases tend to go into separate clusters and SNPs with similar genotype combination distributions tend to go into the same cluster. By ranking the SNPs in a cluster, we only need to investigate the interactions of those SNPs ranking top in different clusters. To do the clustering, the distance of an SNP, $ x_{i} $, to a cluster, $ C $, is defined as

\begin{eqnarray}
Dist(x_{i}, C) = \dfrac{1}{|C|} \sum_{x_{j}\in C} Dist(x_{i}, x_{j})
\label{equ:sam1}
\end{eqnarray}

where $ |C| $ is the number of SNPs in that cluster. In the implementation of JS-MA, we use $ k $-means clustering. There is no geometric coordinate of each SNP in our distance measure, so it is inexplicit to calculate the centroid of each cluster by averaging the sum of coordinates of all SNPs. Instead of doing so, we define the centroid of a cluster as the SNP with the smallest amount of distance to the rest of SNPs in the same cluster. To avoid emulating all pairs of SNPs to obtain the centroid, we propose a heuristic routine, $ Centralizing $, to reduce the time complexity further. Assumption is needed to apply $ Centralizing $ that the SNPs are around particular centroid, and the probability of an SNP showing near the centroid is larger than the possibility of an SNP showing far away the centroid. We call this data as the centred data. The intuition of this heuristic routine to locate the centroid is summarized in Conjecture~\ref{sam:the}. 

\begin{conj}
\label{sam:the}
For a centred data in any dimensional space, given an arbitrary point $ x_{a} $, the distance, $ Dist $, between $ x_{a} $ and the rest points is considered as a random variable. The distance between the centroid $ x_{0} $ and  $ x_{a} $, $ Dist(x_{a},x_{0}) $ is indicating the last mode (if exists) for the probability distribution of $ Dist $.
\end{conj}

Here we do not give the formal proof for the Conjecture~\ref{sam:the} since we do not need very accurate clustering results, and a rough estimation to the $ k $-means clustering will be good enough for the downstream analysis. But we provide the proof to Conjecture~\ref{sam:the} when data points are in 2-dimension space as follows:

\begin{figure}[h!]
\centering
\includegraphics[width=0.6\linewidth]{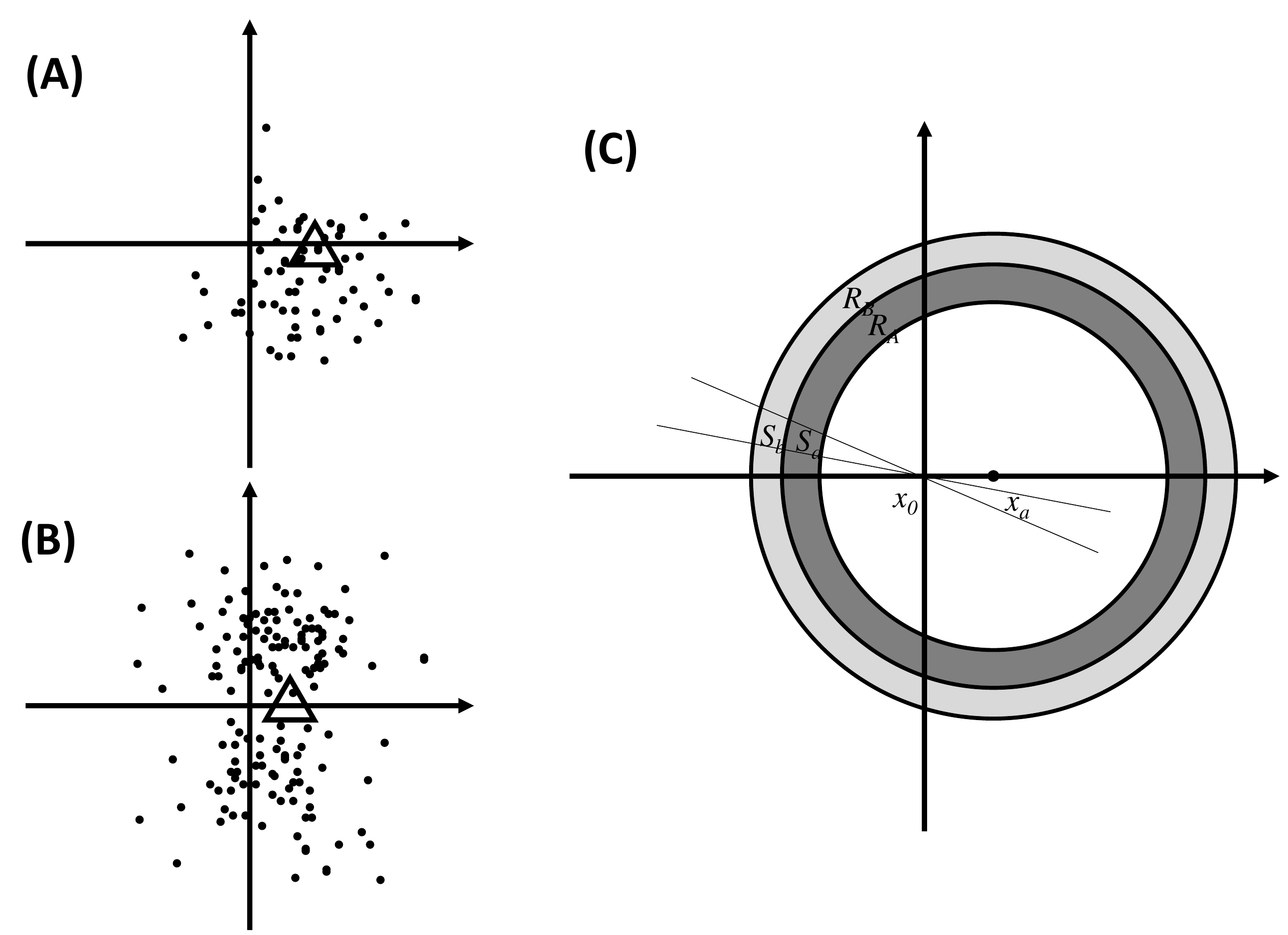}
\caption{Illustration for a centred data in 2-dimensional space.}
\label{fig:sam1}
\end{figure}

\begin{proof} As shown in Figure~\ref{fig:sam1}C, there are two rings, $ R_{A}, R_{B} $, whose area amounts are equal. If there are only one cluster, like Figure~\ref{fig:sam1}A, we can use one ring to cover it. If there are more than one cluster, like two clusters in Figure~\ref{fig:sam1}B, we can still use one ring to cover them. And the centres for both cases are marked by the triangles. So we focus on Figure~\ref{fig:sam1}C that $ R_{A}$ and $ R_{B} $ have the same centroid, $ x_{a} $, and $ x_{0} $ is also inside these two rings. Two lines can be drawn to cut out two pieces, $ s_{a}, s_{b} $, as illustrated in the figure, where the areas of $ s_{a}$ and $ s_{b} $ are equal. Obviously, the possibility of points falling in $ s_{a} $ is larger than the possibility of points falling in $ s_{b} $ because of the centred data assumption. We can cut the rest of $ R_{A}$ and $ R_{B} $ in the same way and we can gain that the possibility of points falling in $ R_{a} $ is larger than the possibility of points falling in $ R_{b} $. Therefore, as the distance getting larger than $ Dist(x_{a},x_{0}) $, the possibility to have a point is decreasing. Based on the mode definition, $ Dist(x_{a},x_{0}) $ is indicating the last mode for the probability distribution of $ Dist $.
\end{proof}

Based on the centred data assumption and Conjecture~\ref{sam:the}, $ Centralizing $ first (1) selects an SNP, $ x_{b} $ randomly and calculates the distances from $ x_{b} $ to the rest of SNPs. It then (2) sorts all SNPs according to the magnitude of distances to $ x_{b} $, and next puts the SNPs into bins with fixed width, like using the histogram to represent graphically the distribution of numerical data. Along the increasing of these distances, it (3) marks the bin in which the number of SNPs starts to decrease as threshold $ \tau $ to filter out SNPs with larger distance to $ x_{b} $. Only those SNPs with distances larger than $ \tau $ will be selected as a new set and $ Centralizing $ reapplies step (1), (2) and (3) until the candidate SNPs for centroid is less than a predefined number. We use 100 for simulation with 1000 SNPs and 2000 for real data with half million SNPs. Based on the centroid candidate, JS-MA takes the $ k $-means clustering to find the centroids and assigned SNPs to those centroids. After clustering, top $ f \times k $ candidates from all clusters based on their ranking scores. Here, $ f $ is a user defined number. A candidate in a cluster is the SNP whose genotype combination frequencies show high dissimilarity between any two groups, in other words, far away from the elements in the other clusters. We define the ranking score as follows

\begin{eqnarray}
Score(x) = \sum_{x\in C_{j}, j\neq i}Dist(x, C_{i})
\label{equ:sam2}
\end{eqnarray}
\noindent
where $ C_{i} $ is the centroid SNP of $ i $-th cluster.

\subsection{Stepwise Evaluation of Interaction}
\label{eva}
With the candidate SNPs, we apply the $ \chi^{2} $ statistic and the conditional $ \chi^{2} $ test similar to~\cite{guo2015dam} to measure the significance for a module of SNPs. Let $ \mathbb{A} = (x_{1}, x_{2}, \ldots, x_{d}: T) $ denote an SNP module $ \mathbb{A} $ with $ d $ SNPs of epistasis type $ T $. We use $ \chi^{2}(x_{1}, x_{2}, \ldots, x_{d}: T) $ to denote the $ \chi^{2} $ statistic of $ \mathbb{A} $ and $ \chi^{2}(x_{1}, x_{2}, \ldots, x_{d}|x_{c_{1}}, x_{c_{2}}, \ldots, x_{c_{d'}}: T) $ as  the conditional $ \chi^{2} $ statistic by given a subset of $ \mathbb{A}' $, $ (x_{c_{1}}, x_{c_{2}}, \ldots, x_{c_{d'}}) $ with $ d' $ SNPs. The $ \chi^{2} $ statistic can be calculated as

\begin{eqnarray}
 \chi^{2}(x_{1}, x_{2}, \ldots, x_{d}: T) = \sum_{i=1}^{|S_{T}|}\sum_{s=1}^{3^{d}}\dfrac{(n_{i, s}-e_{i, s})^{2}}{e_{i, s}}
\label{equ:8}
\end{eqnarray}

\noindent
where $ n_{i, s} $ is the frequency of $ s $-th genotype combination in $ i $-th disjoint set of the epistasis type $ T $, $ e_{i, s} $ is the corresponding expected frequency, and $ S_{T} $ the disjoint set of $ L $ groups. The degrees of freedom for Equation~\ref{equ:8} is $ (|S_{T}| - 1)\cdot(3^{d} - 1) $.

The conditional independent test based on the $ \chi^{2} $ statistic is defined as follows

\begin{eqnarray}
 \chi^{2}(x_{1},\ldots, x_{d}|x_{c_{1}}, \ldots, x_{c_{d'}}: T) = \notag \\ \sum_{\iota=1}^{3^{d'}} \sum_{i=1}^{|S_{T}|}\sum_{s=1}^{3^{d-d'}}\dfrac{(n_{i, s}^{(\iota)}-e_{i, s}^{(\iota)})^{2}}{e_{i, s}^{(\iota)}}
\label{equ:dam9}
\end{eqnarray}

\noindent
where we calculate $ \chi^{2} $ statistic for $ \mathbb{A} - \mathbb{A}' $ separately for the given genotype combinations of $ \mathbb{A}' $. The degrees of freedom for Equation~\ref{equ:dam9} is $ 3^{d'}\cdot(|S_{T}| - 1)\cdot(3^{d-d'} - 1) $. We treat those SNPs as redundant SNPs when they are conditional independent by giving a subset of the SNP module. To avoid these redundant SNPs, we use the same definition of the compact epistatic interaction to select qualified SNP modules~\cite{guo2015dam} as follows:

\begin{mydef}
an SNPs module $ \mathbb{A} = (x_{1}, x_{2}, \ldots, x_{d}: T) $ is considered as a significant compact interaction by giving a significant level $ \alpha_{d} $, if it meets the following three conditions:\\
(1) the p-value of $ \chi^{2}(x_{1}, x_{2}, \ldots, x_{d}: T) \leq \alpha_{d} $; \\
(2) the p-value of $ \chi^{2}(x_{1}, x_{2}, \ldots, x_{d}: T) = $ the minimum p-value of $ \chi^{2}(x_{1}, x_{2}, \ldots, x_{d}: T)$;\\
(3) the p-value of $ \chi^{2}(x_{1}, x_{2}, \ldots, x_{d}|x_{c_{1}}, x_{c_{2}}, \ldots, x_{c_{d'}}: T) \leq \alpha_{d} $ for $ \forall\mathbb{A}' = (x_{c_{1}}, x_{c_{2}}, \ldots, x_{c_{d'}}: T)$ whose p-value $ \leq \alpha_{d'} $.
\label{def:dam}
\end{mydef}

Based on the Definition~\ref{def:dam}, we develop a stepwise algorithm to search for top-$ f $ significant $ d $-locus significant compact interactions, where the searching space only includes the SNP markers generated by the last clustering step. We assume that one SNP can only participate in one significant interaction with one type. We first searches all the modules with just one SNP based on Definition~\ref{def:dam}, then we recursively tests all the possible combinations by setting the module size with one more SNP. For the SNPs reported as jointly contributing to the disease risk, we calculate the $ p $-value under different types and use the conditional test if part of SNPs already reported as significant. All SNPs with significant marginal associations after a Bonferroni correction are reported in a list $ \mathbb{L} $. The algorithm recursively searches the interaction space with larger module size until $ d $ reaches a user pre-set value. We add all novel $ d $-way interactions (i.e. none of the SNPs in the module has been reported earlier) that are significant to $ \mathbb{L} $ after the Bonferroni correction for $ B\cdot{{M}\choose{d}} $ tests. For the interactions whose subsets have been reported as significant before, we use the conditional independent test, and put the interaction in $ \mathbb{L} $ if it is still significant after Bonferroni correction of $ B\cdot{{M}\choose{d}}\cdot{{d}\choose{d'}} $ tests. We also apply a distance constraint that the physical distance between two SNPs in a multi-locus module should be at least 1Mb. This constraint is used to avoid associations that might be due to the linkage disequilibrium effects~\cite{introEps2}.

\subsection{Algorithm}

The details of the JS-MA algorithm is shown in Algorithm~\ref{alg:samalg1} consisting three parts: clustering, candidates ranking, and stepwise evaluation. The convergence of $ k $-means clustering is very fast on the simulated data. Usually we set the number of iteration to 50 which is large enough to get a good estimation of the SNPs' belongings. Inside the $ k $-means clustering, the distance between each SNP and the centroid is computed according to Equation~\ref{equ:sam1} and the $ Centralizing $ procedure is employed to update the centroids. In the second part, all SNPs are ranked based on Equation~\ref{equ:sam2} and inserted into a size-limited descending list to select promising candidates. In the last part, the $ \chi^{2} $ statistic and the conditional $ \chi^{2} $ test are used to find the epistatic interactions.

\begin{algorithm}[!h]
\footnotesize
\caption{The JS-MA Algorithm}
\label{alg:samalg1}
\begin{algorithmic}[1]
\REQUIRE{An $ N \times (M+1) $ matrix}

\STATE Read $ N \times (M+1) $ matrix file\;
\STATE Apply the procedure $ Centralizing $ to obtain centroid candidates\;
\STATE Initialize $ n_{iteration} $\;
\STATE Randomly select $ k $ SNPs from the centroid candidates as centroid\;
\FOR{ $ i $ in $ n_{iteration} $ }
	\FOR{each SNP $ x $ in centroid candidates}
		\STATE Calculate $ DIST(x, C) $ \;
		\STATE Assign $ x $ to a cluster \;
	\ENDFOR
	\STATE Assign all SNPs to a cluster \;
\ENDFOR

\STATE Initialize descending list $ \mathbb{L} $ with length $ fk $\;
\FOR{each SNP $ x $}
	\STATE Calculate $ Score(x) $\;
	\STATE Place $ x $ into $ \mathbb{L} $ if $ Score(x) $ is among top $ fk $ SNPs \;
\ENDFOR

\STATE Stepwise evaluate all possible SNP modules based on $ \mathbb{L} $ \;
\end{algorithmic}

\end{algorithm}

\section{Results and Discussion}

We first introduce the definitions of 4 two-locus and 3 three-locus multi-disease models, the power metric, and then evaluate the effectiveness of JS-MA comparing to DAM. Before showing the experiments on the simulated datasets, we present the false positive rate of JS-MA on the null simulation for testing type I errors. We also present results of JS-MA on two real GWAS datasets, i.e. Rheumatoid Arthritis (RA) and Type 1 Diabetes (T1D). Interactions detected by JS-MA from different orders demonstrate a great number of novel, potentially disease-related genetic factors.

\subsection{Experimental design}
\paragraph{Data simulation} To evaluate the performance of JS-MA, we perform extensive simulation experiments using 4 two-locus disease models (Model 1-4) and 3 three-locus models (Model 5-7) with three groups, including 2 case groups and 1 control group. The odds tables describing these 7 models are in the Supplementary Material. For the 4 two-locus disease models with marginal effects, we take the same parameters as those in~\cite{BOOST}~\cite{DCHE}, namely, $ h^{2} $=0.03 for Model 1, $ h^{2} $=0.02 for Models 2, 3 and 4 and $ p(D) $=0.1 for all 4 models. Minor allele frequencies ($ maf $) are set to three levels: $ \left\lbrace 0.1, 0.2, 0.4 \right\rbrace  $. For the 3 three-locus models, Model 5 shows the  multiplicative effect between and within loci, Model 6 shows the multiplicative effect between loci, and Model 7 shows the threshold effect. We set $ h^{2} $=0.03 and $ p(D) $=0.1 for Model 5, 6 and 7. The solved parameters $ \alpha $ and $ \theta $ under different settings are listed in the Supplementary Material. The genotypes of unassociated SNP are generated by the same procedure used in previous studies~\cite{DCHE} with $ mafs $ sampled from $ \left[ 0.05, 0.5 \right] $. The six settings of epistasis types are shown in Table~\ref{tab:damsetting}. As introduced in the section~\ref{note}, ET 1 indicates the loci only contributing to the first case status, ET 2 indicates the loci only contributing to the second case status, ET 3 indicates the loci showing equal effect on both case groups, and ET 4 indicates the loci showing different effects for each case group. By given an $ maf $, we generate 100 replicas for each setting. Note that there are only 3 epistatic interactions (ET 1, 2, and 3) in Setting 1 and 4, because the combination of 3 two-locus models is unsolvable when $ maf= 0.1 $. Each simulated replica contains $ M = 1000 $ SNPs. The sizes of two case groups and one control group are set to $ (500, 500, 1000) $ or $ (1000, 1000, 2000) $.

\begin{table}[h]
\centering
\caption{Six settings with 4 epistasis types whose effect sizes based on 7 disease models.}
\begin{tabular}{ccccc}
\hline
Epistasis Type (ET) & 1       & 2       & 3       & 4            \\ \hline \hline
Setting 1 & Model 1 & Model 1 & Model 1 & Model 1 \& 2 \\ 
Setting 2 & Model 2 & Model 2 & Model 2 & Model 2 \& 3 \\ 
Setting 3 & Model 3 & Model 3 & Model 3 & Model 3 \& 4 \\ 
Setting 4 & Model 4 & Model 4 & Model 4 & Model 4 \& 1 \\ 
Setting 5 & -- & -- & -- & Model 5 \& 7 \\ 
Setting 6 & -- & -- & -- & Model 6 \& 7 \\ \hline
\end{tabular}
\label{tab:damsetting}
\end{table}

\paragraph{Statistical power} In the evaluation of performances on simulated data, 100 datasets are generated for each setting. The measure of discrimination power is defined as the fraction of 100 datasets on which the ground-truth associations are identified as compact and significant by the association mapping algorithms.

\subsection{Null Simulation to Test Type I Errors}
\begin{figure}[h]
\centering
\includegraphics[width=\linewidth]{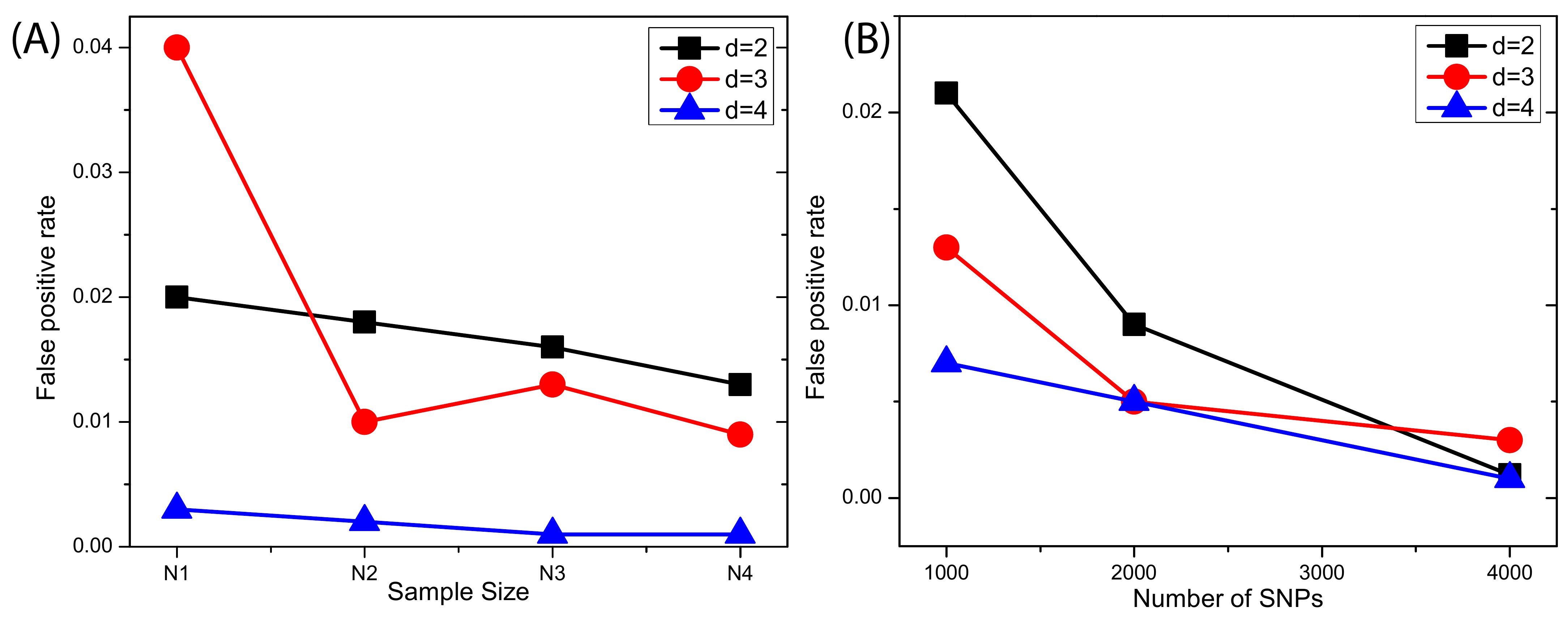}
\caption{False positive rates of JS-MA under null simulation. The plots in (A) and (B) show the false positive rates for different $ d $s when group sizes and the numbers of SNP vary.}
\label{fig:samfdr}
\end{figure}

Here we examine the false positive rate for varied sizes of interaction, \textit{i.e.} $ d = 2, 3, 4 $. We generate 1000 null data sets for 6 settings, respectively. Specifically, we fix the number of SNP to 1000 and vary the number of samples in each group. The first four settings regarding the numbers of the individuals are $ N1 = \left\lbrace 200, 200, 400 \right\rbrace $, $ N2 = \left\lbrace 400, 400, 800 \right\rbrace $, $ N3 = \left\lbrace 800, 800, 1600 \right\rbrace $, and $ N4 = \left\lbrace 1600, 1600, 3200 \right\rbrace $, where the first two numbers indicate the sizes of two case groups, and the last number is the size of the control group. For the last two settings, using $ N4 $ as the groups' size, we increase the number of SNP to 2000 and 4000. All of the SNPs are generated independently, with $ maf $ uniformly distributed in $ [0.05, 0.5] $. Note that we set the significance level to 0.1 and also apply the Bonferroni correction for multiple comparisons. The degree of freedom for Pearson's $ \chi^{2} $ test is $ df = (|T| - 1)(|G| - 1) $, where $ T $ denotes the type and $ G $ is the set of genotype given the SNP module. The degree of freedom for conditional $ \chi^{2} $ test is $ |G'|(|T| - 1)(|G / G'| - 1) $, where $ G' $ is the set of genotype given the subset of SNP module and $ G/G' $ is the set of genotype for the rest SNPs. The results are shown in Figure~\ref{fig:samfdr} in which JS-MA has lower type I error rate than the given significant level.

\subsection{Simulation Experiments on Two-locus Models}

\begin{figure}[h!]
\centering
\includegraphics[width=0.8\linewidth]{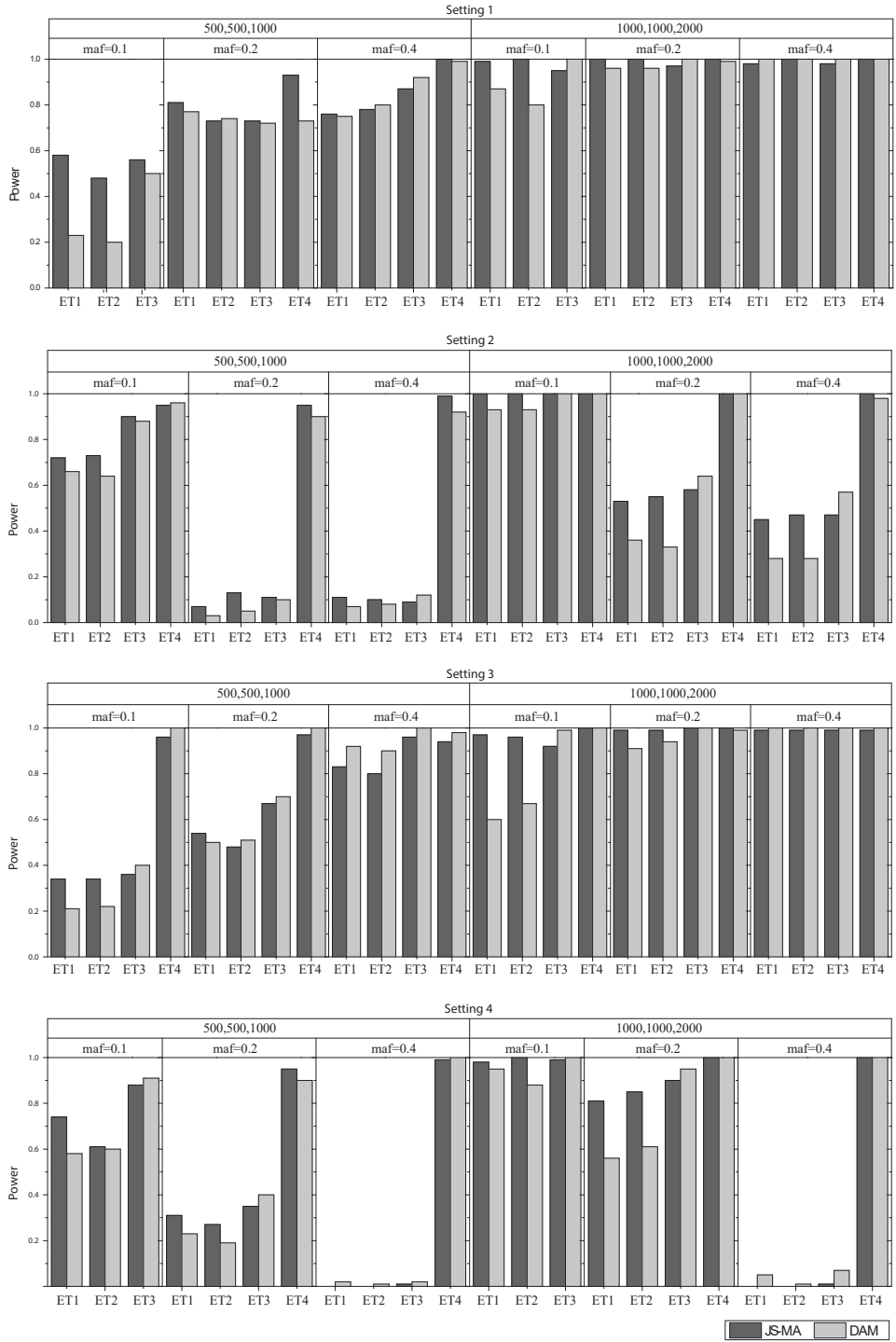}
\caption{Performance comparison between JS-AM and DAM on the simulated settings 1, 2, 3 and 4. Note that the combinations of model 1 with the rest 3 two-locus models have no mathematical solution when $ maf = 0.1 $.}
\label{fig:Joint}
\end{figure}

We test the performance of JS-MA and DAM on the datasets with 1000 SNPs. The test results are illustrated in Figure~\ref{fig:Joint}. As expected, both methods have higher power when the sample size increases from (500, 500, 1000) to (1000, 1000, 2000). For setting 1 and 3, the power of both algorithms increases when the $ maf $s of the disease associated SNPs vary from 0.1 to 0.4. The trends are unclear for Setting 2 and 4. From the distributions of main effects and interaction effects illustrated in~\cite{Wan2}, the trends for setting 2 and 4 are caused by the relative lower main effects. If the main effects are too low, all algorithms will lose powers without the brute-forth enumeration; but the power of JS-MA is higher except a few cases where the power is comparable with DAM. For these 96 simulated associations, JS-MA outperforms DAM in 46 parameter combinations, while they are comparable in the remaining associations. For a more intuitive comparison, we adopt the concept~\cite{DCHE}, the overall quality $ q=n_{correct}/n_{total} $, where $  n_{correct} $ is the number of datasets where programs successfully detect the ground-truth interactions and $ n_{total} $ is the total number of datasets. Following this definition, The overall quality of JS-MA and DAM are 0.595 and 0.564 for the sample size of $ \left( 500, 500, 1000 \right)  $, and 0.853 and 0.805 for the sample size of $ \left( 1000, 1000, 2000 \right) $, respectively. We can find that along the increasing of the sample size, there about 3 to 5 percent overall quality difference between JS-MA and DAM.

\subsection{Simulation Experiments on Three-locus Models}

\begin{figure}[h!]
\centering
\includegraphics[width=0.6\linewidth]{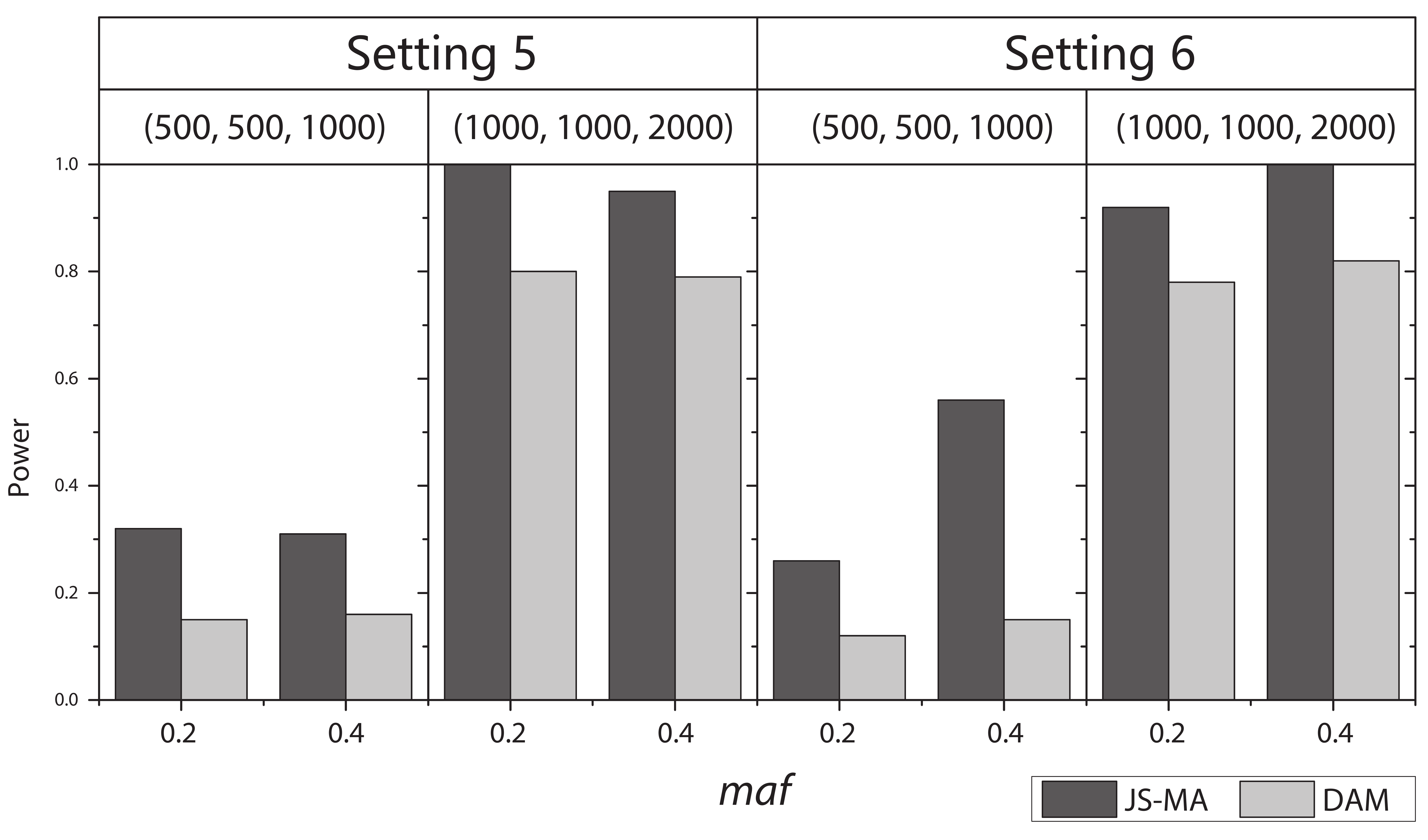}
\caption{Performance comparison between JS-AM and DAM on the simulated settings 5 and 6.}
\label{fig:Three}
\end{figure}

The experimental results on Setting 5 and 6 are shown in Figure~\ref{fig:Three}, respectively. Only associations of epistatsis type 4 are embedded into the datasets. The reason we do not show the experiments for other epistatsis types is that model 5, 6 and 7 are extensions to model 1. Our goal is to exhibit the capability of JS-AM to map those SNP modules having distinguish effects on different diseases. Therefore, epistasis type 4 is more suitable for this simulation experiment. The results of setting 5 and 6 in Figure~\ref{fig:Three} show similar trend that power increases when $ maf $ is larger. In both scenarios, JS-MA shows stronger power and significantly outperform DAM. Using the same overall quality measurement, JS-MA obtains 0.645 and 0.685 for setting 5 and 6; while DAM obtains 0.475 and 0.467 for setting 5 and 6.

\subsection{Computation Time}

From a practical point of view, a challenging bottleneck of mapping multiple loci epistatic interactions in genome-wide case-control studies is the computational efficiency. Traditional two-locus epistatic interaction detection usually takes several days to finish the analysis of a couple millions SNPs on a standard desktop~\cite{BOOST}. We tested the running time of JS-MA and DAM on our desktop computer, which comes with Windows 8 OS, Intel i5-3337 CPU of 1.8 GHz, and 8 GB memory. The results are shown in Table~\ref{tab:timeSD}. We can see that JS-MA is at least 7 times faster than DAM in all 3 scenarios, and the time used by JS-MA is greatly reduced when the number of SNPs increases.

\begin{table}[h]
\centering
\caption{Time Comparison of JS-MA and DAM (in seconds.)}
\label{tab:timeSD}
\begin{tabular}{lll}
\hline
Data size         & JS-MA        & DAM         \\ \hline
N=6,000, M=1,000  & 8.6s     & 131.4s    \\
N=6,000, M=5,000  & 60.4s    & 793.2s    \\
N=6,000, M=10,000 & 359.8s, & 2165.1s \\ \hline
\end{tabular}
\end{table}

\subsection{Experiments on The WTCCC Data}

We applied DAM to analyze data from the WTCCC (3999 cases in total and 3004 shared controls) on two common human diseases: Rheumatoid Arthritis (RA), Type 1 Diabetes (T1D), where RA is treated as group 1, T1D is treated as group 2, and control group is group 3. The procedure of quality control is the same as presented in the~\cite{DCHE}. After the SNP filtration, the data set finally has 333,739 high-quality SNPs. JS-MA ran about 2 hours in the machine used for the computation time analysis by setting $ fk = 100 $ with $ k = 10 $ as the number of clusters. JS-MA also reports some novel epistatic interactions. For example, (rs6679677, rs805301) with ET 4 and $ p $-value $ 6.2 \times 10^{-120} $ from the $ \chi^{2} $ test. rs6679677, which is located on Chromosome 1, has been reported to be associated with both RA and T1D~\cite{burton2007genome}. The association between rs6679677 and T1D is actually due to a closely linked, potentially causal variant identified as rs2476601, which is also known as Arg620Trp~\cite{smyth2008ptpn22}. rs805301 is located inside gene BAG6 on Chromosome 6. BAG6 encodes a nuclear protein that forms a complex with E1A binding protein p300 and is required for the response to DNA damage. This SNP module shows different association effects with RA and T1D compared to control group. Another instance is (rs200991, rs11171739) with ET 2 and $ p $-value $ 6.7\times 10^{-26} $ from the $ \chi^{2} $ test. rs200991 is located on Chromosome 6 near the gene, HIST1H2BN, which encodes Histone H2B type 1-N. Histones play a central role in transcription regulation, DNA repair, DNA replication and chromosomal stability. rs11171739 has been reported to be associated with T1D~\cite{burton2007genome}. ET 2 means this SNP module may have some connection associated to RA which has not been found by previous study. JS-MA also reports some 3-locus epistatic interactions. For instance, (rs6679677, rs377763, rs9273363) with ET 2 and $ p $-value $ 1.3\times 10^{-116} $. Both rs377763 and rs9273363 are located on Chromosome 6. rs377763 is near the downstream of the gene NOTCH4, which is found to be associated with multiple sclerosis, a chronic inflammatory disease. rs9273363 is inside the gene HLA-DQA1 which plays a critical role in the immune system. The protein produced from the HLA-DQA1 gene binds to the protein produced from the MHC class II gene, HLA-DQB2. Many studies have reported the MHC region in chromosome 6 with respect to infection, inflammation, autoimmunity, and transplant medicine~\cite{lechler2000hla}\cite{BOOST}\cite{zhang2012high}. A 4-locus interaction found by JS-MA is (rs10924239, rs17432869, rs7610077, rs11098422) with ET 4 and $ p $-value $ 3.9 \times 10^{-106} $. rs10924239 is an intron variant of the gene KIF26B on Chromosome 1. KIF26B is essential for embryonic kidney development. rs17432869 is located on Chromosome 2 and inside the gene LOC105373439 which is an RNA Gene, and is affiliated with the ncRNA class. the rs7610077 is located on Chromosome 3 and inside the gene, SNX4, which encodes a member of the sorting nexin family. rs11098422 is located on Chromosome 4 and inside the gene, NDST3, whose expression impacts the cardiovascular system. Though the validation of relationship of these modules to RA and T1D is beyond the scope of this work, the significant enrichment of some genotype combinations from these modules in both cases implies that they might interact and/or be associated with these two diseases.

\section{Conclusion}
The enormous number of SNPs genotyped in genome-wide case-control studies poses a significant computational challenge in the identification of gene-gene interactions. During the last few years, many computational and statistical tools are developed to finding gene-gene interactions for data with only two groups, \textit{i.e.} case and control groups. Here, we present a novel method, named ``JS-MA'', to address the computation and statistical power issues for multiple diseases GWASs. We have successfully applied our methods to systematic simulation and also analyzed two datasets from WTCCC. Our experimental results on both simulated and real data demonstrate that our methods are capable of detecting high-order epistatic interactions for multiple diseases at the genome-wide scale.
\Acknowledgements{This work was supported by the Molecular Basis of Disease (MBD) program at Georgia State University.}

\InterestConflict


\bibliography{diss}
\bibliographystyle{IEEEtran}







\end{document}